\documentclass[%
  aip, apl,
  amsmath,amssymb,
  reprint,
]{revtex4-2}

\usepackage{graphicx}
\usepackage{dcolumn}
\usepackage{bm}

\usepackage{color}
\usepackage{subcaption}

\usepackage[utf8]{inputenc}
\usepackage[T1]{fontenc}
\usepackage{mathptmx}
\usepackage{etoolbox}
\usepackage{comment}

\usepackage[colorlinks = true]{hyperref}

\hypersetup{colorlinks=true, linkcolor=blue, citecolor=blue, urlcolor=black, breaklinks=true}

\graphicspath{{./figure/}}

\newcommand{\mc}{\mathcal}

\newcommand{\mr}{\mathrm}

\usepackage{algorithm}  
\usepackage{algpseudocode}  

\makeatletter
\def\@email#1#2{%
 \endgroup
 \patchcmd{\titleblock@produce}
  {\frontmatter@RRAPformat}
  {\frontmatter@RRAPformat{\produce@RRAP{*#1\href{mailto:#2}{#2}}}\frontmatter@RRAPformat}
  {}{}
}%
\makeatother
\begin{document}

\preprint{AIP/123-QED}

\title{Practical hybrid PQC-QKD protocols with enhanced security and performance}

\author{Pei Zeng}
\affiliation{Pritzker School of Molecular Engineering, The University of Chicago, Chicago, IL 60637}
\author{Debayan Bandyopadhyay}
\affiliation{Pritzker School of Molecular Engineering, The University of Chicago, Chicago, IL 60637}
\author{Jos\'e A. M\'endez M\'endez}
\affiliation{Pritzker School of Molecular Engineering, The University of Chicago, Chicago, IL 60637}
\author{Nolan Bitner}
\affiliation{Pritzker School of Molecular Engineering, The University of Chicago, Chicago, IL 60637}
\affiliation{Materials Science Division, Argonne National Laboratory, Lemont, IL 60439}
\affiliation{Center for Molecular Engineering, Argonne National Laboratory, Lemont, IL 60439}
\author{Alexander Kolar}
\affiliation{Pritzker School of Molecular Engineering, The University of Chicago, Chicago, IL 60637}
\author{Michael T. Solomon}
\affiliation{Pritzker School of Molecular Engineering, The University of Chicago, Chicago, IL 60637}
\affiliation{Materials Science Division, Argonne National Laboratory, Lemont, IL 60439}
\affiliation{Center for Molecular Engineering, Argonne National Laboratory, Lemont, IL 60439}
\author{Ziyu Ye}
\affiliation{Department of Computer Science, The University of Chicago, Chicago, IL 60637}
\affiliation{SeQure, Chicago, IL 60637}
\author{Filip Rozp\k{e}dek}
\affiliation{College of Information and Computer Sciences, University of Massachusetts Amherst, Amherst, MA 01003}
\author{Tian Zhong}
\affiliation{Pritzker School of Molecular Engineering, The University of Chicago, Chicago, IL 60637}
\author{F. Joseph Heremans}
\affiliation{Pritzker School of Molecular Engineering, The University of Chicago, Chicago, IL 60637}
\affiliation{Materials Science Division, Argonne National Laboratory, Lemont, IL 60439}
\affiliation{Center for Molecular Engineering, Argonne National Laboratory, Lemont, IL 60439}
\author{David D. Awschalom}
\affiliation{Pritzker School of Molecular Engineering, The University of Chicago, Chicago, IL 60637}
\affiliation{Materials Science Division, Argonne National Laboratory, Lemont, IL 60439}
\affiliation{Center for Molecular Engineering, Argonne National Laboratory, Lemont, IL 60439}
\affiliation{Department of Physics, University of Chicago, Chicago, IL 60637}
\author{Liang Jiang}
\affiliation{Pritzker School of Molecular Engineering, The University of Chicago, Chicago, IL 60637}
\author{Junyu Liu}
\affiliation{Pritzker School of Molecular Engineering, The University of Chicago, Chicago, IL 60637}
\affiliation{Department of Computer Science, The University of Chicago, Chicago, IL 60637}
\affiliation{SeQure, Chicago, IL 60637}
\affiliation{Department of Computer Science, The University of Pittsburgh, Pittsburgh, PA 15260}
\email{junyuliucaltech@gmail.com,liangjiang@uchicago.edu}

\date{\today}

\begin{abstract}
    Quantum resistance is vital for emerging cryptographic systems as quantum technologies continue to advance towards large-scale, fault-tolerant quantum computers.
    Resistance may be offered by quantum key distribution (QKD), which provides information-theoretic security using quantum states of photons, but may be limited by transmission loss at long distances.
    An alternative approach uses classical means and is conjectured to be resistant to quantum attacks\textemdash so-called post-quantum cryptography (PQC)\textemdash but it is yet to be rigorously proven, and its current implementations are computationally expensive.
    To overcome the security and performance challenges present in each, here we develop hybrid protocols by which QKD and PQC inter-operate within a joint quantum-classical network.
    In particular, we consider different hybrid designs that may offer enhanced speed and/or security over the individual performance of either approach.
    Furthermore, we present a method for analyzing the security and performance of hybrid protocols in key distribution networks.
    Our hybrid approach paves the way for joint quantum-classical communication networks, which leverage the advantages of both QKD and PQC and can be tailored to the requirements of various practical networks.
\end{abstract}

\maketitle

In recent years, we have seen a rapid development of quantum information science and technologies. This necessitates the construction of secure communication networks and cryptographic systems that are able to withstand attacks from future quantum computers. Quantum key distribution (QKD) is a leading approach developed to address this need~\cite{bennett1984quantum,ekert1992quantum,scarani2009security}. In QKD protocols, the communication parties transmit encoded quantum states and perform quantum measurements to distribute symmetric keys, with information-theoretic security based on the fundamental principles of quantum mechanics~\cite{gisin_quantum_2002,zapatero2023}. Although there are many successful examples in academia and industry of implementing metropolitan, intercity, or even global QKD networks~\cite{peev2009secoqc,sasaki2011field,tang2016measurement,chen2021integrated}, the long-distance performance of fiber-based QKD technologies is significantly limited by the exponential decay of key rates over increasing distances~\cite{diamanti_practical_2016}. Additionally, the security of practical QKD systems might be affected by the imperfections in the light sources and measurement devices~\cite{xu2020secure}, which require extra counter-measure designs to close the loopholes and hence introduce extra device complexities~\cite{scarani2009security,xu2020secure}.

Post-quantum cryptography (PQC) is another promising technique to provide quantum resistance~\cite{bernstein2017post}, using only classical methods. Unlike traditional cryptography based on classically hard problems like factoring, elliptic curves, or discrete logarithms that are nonetheless vulnerable to efficient quantum algorithms~\cite{shor1999polynomial}, PQC leverages problems conjectured to be hard even for quantum computers~\cite{bernstein2017post}. 
Due to its classical nature, PQC can be readily deployed in current cryptographic systems with existing hardware, and its communication rate is not limited by transmission distance. In fact, the National Institute of Standards and Technology (NIST) has already called for the standardization of certain PQC protocols~\cite{nist}, including CRYSTALS-Kyber~\footnote{In the final NIST standard~\cite{nist}, Kyber has been modified and renamed as the Module-Lattice-Based Key-Encapsulation-Mechanism (ML-KEM). In this work, we will not differentiate between these two protocols.}, a key sharing algorithm based on a variant of the lattice problem known as Learning with Errors~\cite{bos2018crystals}. On the other hand, deployment of PQC protocols still faces significant challenges. Firstly, the security of PQC has not been conclusively established. Evolving research on algorithms to break various PQC schemes~\cite{castryck2022efficient} continues to be met with varying success~\cite{eldar2016efficient,chen2024quantum}. Secondly, implementations of PQC algorithms demand considerable computational effort. As the resulting PQC key rates depend heavily on the computational power of the users, performance is less than ideal when compared to the existing mature cryptographic infrastructure deployed throughout modern communication networks. This limits its commercial viability with personal computers and the scope of real-world use, although purpose-built chips could alleviate this concern in the future.

In light of these limitations, PQC has been utilized to improve aspects of the classical exchange required for QKD, including authentication~\cite{yang2021all} and information reconciliation~\cite{djordjevic2020joint}. Recent work has also begun investigating specific single-link joint PQC-QKD protocols and cryptographically evaluating the operational issues of link security~\cite{dowling2020many,garms2024experimental}. In this work, we explore the construction of a composite symmetric key distribution system that integrates PQC with QKD, leveraging the advantages of both. Our work analyzes multiple combinations of these key distribution mechanisms across network elements, which can then be concatenated to act as an overall symmetric key sharing scheme between two users in a network. For any such scheme, our analysis can enumerate the vulnerabilities and calculate the end-to-end secure key generation rates. This analysis could be used to engineer networks with optimal security and performance.

\begin{figure}[t]
    \includegraphics[width=0.8\columnwidth]{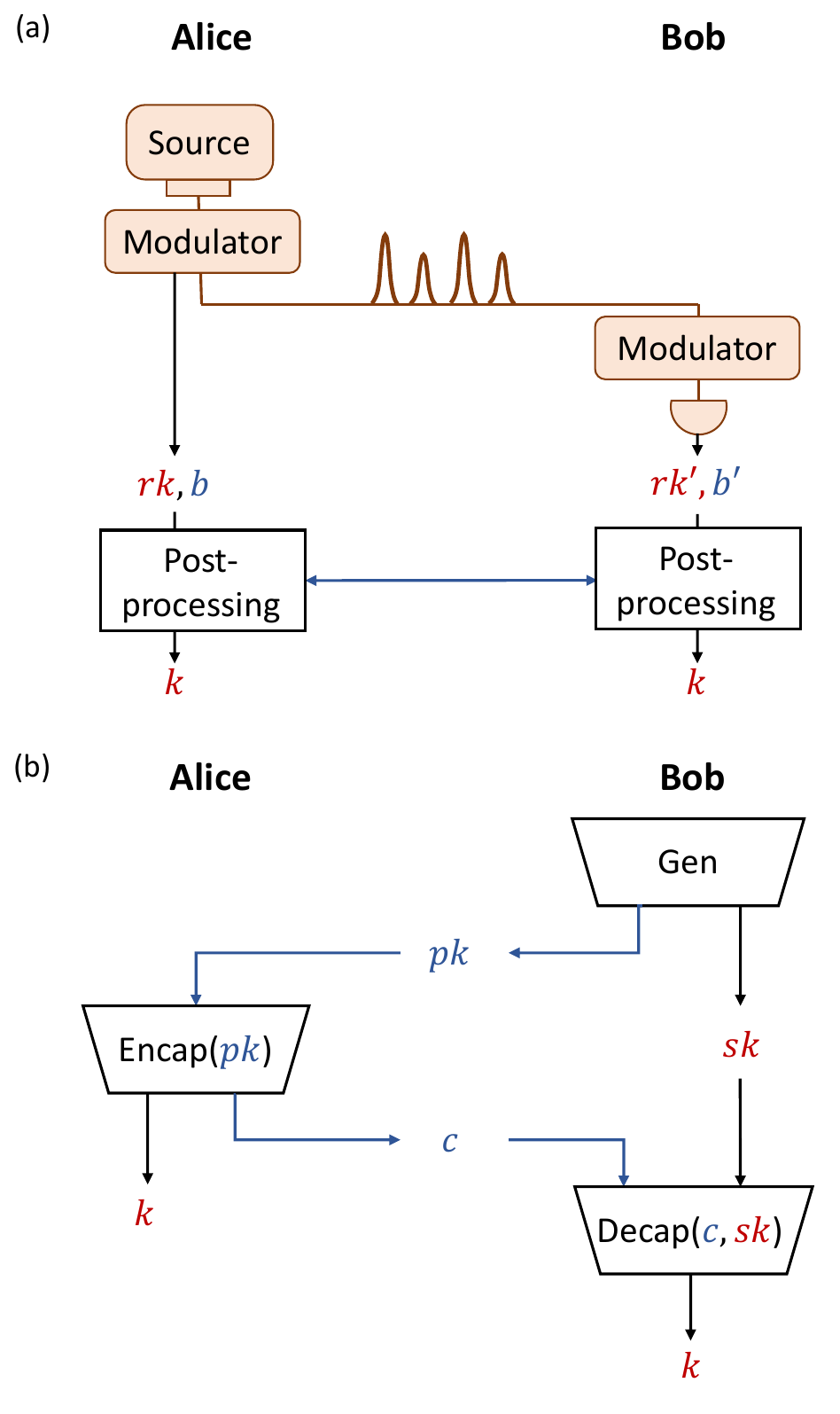}
    \caption{Illustration of two symmetric key distribution protocols used in our framework. (a) Prepare-and-measure quantum key distribution (QKD) protocols. (b)  Key-encapsulation mechanism based on a post-quantum cryptography (PQC) system. }
    \label{fig:PQCQKDprotocol}
\end{figure}

One major component we use to construct the symmetric key distribution network is prepare-and measure QKD~\cite{scarani2009security,xu2020secure}, shown in Fig.~\ref{fig:PQCQKDprotocol}(a). In this protocol Alice generates a random raw key $rk$, which she encodes into quantum states that she transmits to Bob. She records classical information $b$ about the encoding such as the basis used. Bob receives the quantum states and measures them to obtain $rk'$ and classical information $b'$ related to the measurement procedure. Alice and Bob then announce $b$ and $b'$ as well as a subset of $rk$ and $rk'$ and perform classical post-processing, including security parameter estimation, error correction, and privacy amplification, to generate the final symmetric key bits $k$ from the raw key bits $rk(rk')$.

The other main component we consider is a key encapsulation mechanism (KEM) based on a post-quantum public-key cryptography system. KEM~\cite{dent2003designer} is a widely-used classical cryptographic method for distributing symmetric keys using public-key encryption (PKE) (see Ref. \cite{katz2020introduction} for a general introduction). The basic idea of KEM is to use PKE to distribute a random message from Alice to Bob that they can keep as symmetric key. As is shown in Fig.~\ref{fig:PQCQKDprotocol}(b), a typical process of KEM involves three steps: key generation, encapsulation, and decapsulation. In the key generation step, Bob runs PKE to generate a public key $pk$ and a private key $sk$ used for encryption and decryption. He then announces $pk$ to Alice. In the encapsulation step, Alice uses $pk$ to simultaneously generate and encrypt a random message $k$ to its ciphertext $c$. She then announces $c$ to Bob. Finally, in the decapsulation step, Bob uses $sk$ to decrypt $c$ and obtains $k$. They then store $k$ for later use as a symmetric key.

The motivation for using a Key Encapsulation Mechanism (KEM) instead of directly relying on Public-Key Encryption (PKE) is to enhance the practical security of the PKE system. 
In modern security models for public-key systems, the eavesdropper (hereafter "Eve") is often allowed to query the encryption and/or decryption schemes to study their behavior.
In a chosen-plaintext attack (CPA), Eve can select arbitrary plaintexts and obtain the corresponding ciphertexts to analyze the encryption scheme, aiming to deduce information about the encryption key. A stronger attack, a chosen-ciphertext attack (CCA), allows Eve to choose ciphertexts and learn the decrypted plaintexts, which can help compromise the encryption system by exploiting this decryption information. 

While many typical PKE systems can prevent a CPA-capable Eve from learning the symmetric key bits, it is often difficult to prove that these systems are also secure against a CCA-capable Eve~\cite{katz2020introduction,cramer2003design}. 
By using KEM, in which the deterministic message is replaced with random bits, it becomes difficult for Eve to learn the key bits, even under a CCA.
While KEMs were traditionally developed with classical attackers in mind, any realistic eavesdropper should now be assumed to have access to scalable quantum computation. Thus, in our discussion of hybrid protocols, we assert that any KEM must be based on an underlying post-quantum cryptographic system. For example, the PKE provided by CRYSTALS-Kyber is guaranteed security under a CPA if the module-learning-with-error problem~\cite{bos2018crystals} is hard for quantum computers. By introducing additional randomness through the Fujisaki-Okamoto transformation~\cite{fujisaki1999secure,hofheinz2017modular}, Kyber provides a CCA-secure KEM scheme under the assumption of a quantum random oracle model. 

\begin{figure*}[t]
    \includegraphics[width=1\textwidth]{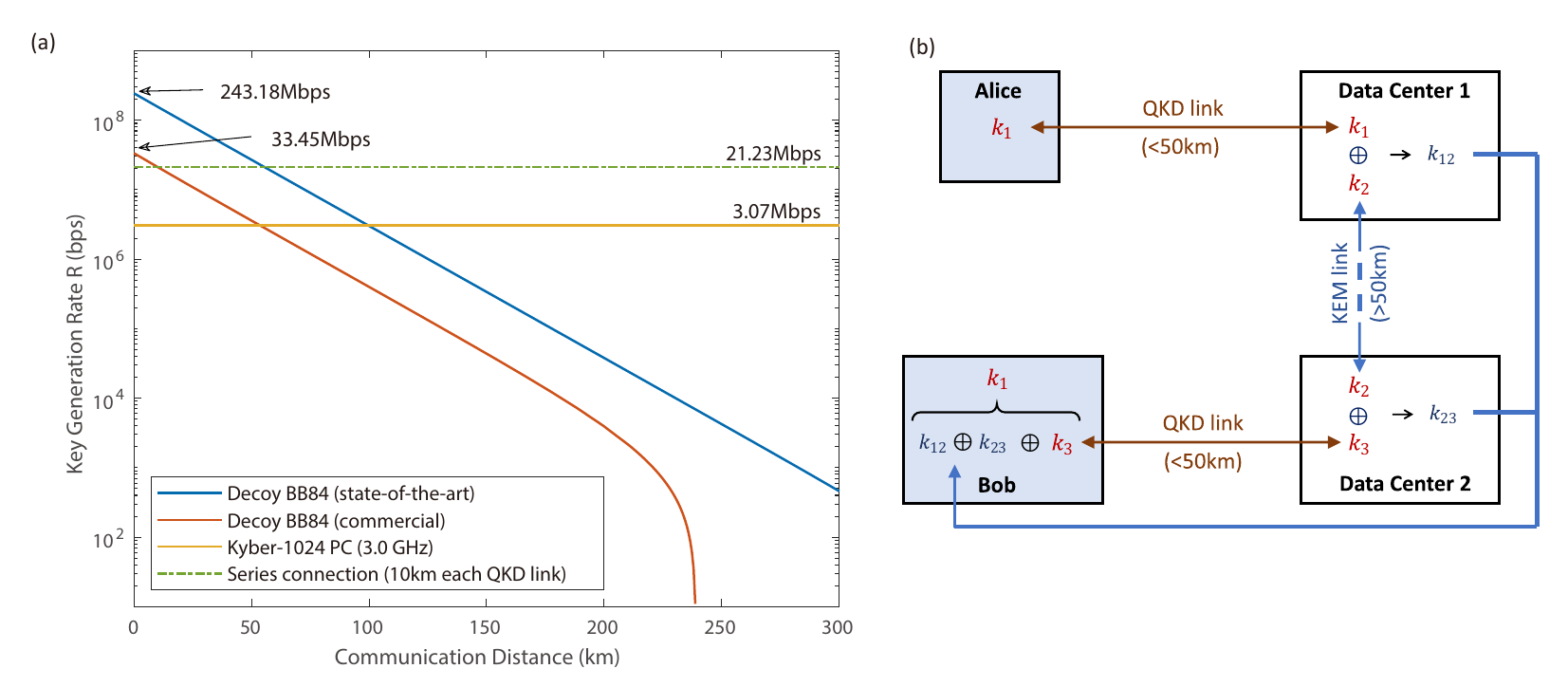}
    \caption{(a) Performance comparison of different symmetric key distribution protocols with respect to communication distance.
    (b) Design of a series-connection protocol where the end-user performance is higher than the bare usage of KEM or QKD without relay nodes. An example of the performance of this protocol for QKD links of length 10 km is plotted as a dashed green line in panel a.}
    \label{fig:series}
\end{figure*}
One primary motivation to combine KEM and QKD into a single protocol is to enhance the key generation rate over long distances. In Fig.~\ref{fig:series}a, we compare the key generation speed of KEM and point-to-point QKD with respect to communication distance. For the KEM performance estimation, we consider users running Kyber-1024~\cite{kyber,bos2018crystals} on their personal computers with 3.0 GHz clock frequency. For the QKD performance estimation, we assume a commercial fiber link between users with a loss of $0.19$ dB/km. For the commercial and state-of-the-art QKD performance, we mainly consider the parameters in Yuan~et~al.~\cite{yuan2018T12} and Li~et~al.~\cite{li2023high}, respectively. The details of the performance estimation can be found in the supplementary materials~\cite{supplementary}. For concreteness, we assume that end users do not employ signal multiplexing for QKD or parallelized computation for PQC.

We observe that QKD outperforms KEM when computational power is limited, particularly for short communication distances (less than 50 km). This advantage arises because, at shorter distances, the key generation rate of QKD is primarily constrained by the clock rate of the source and the dead time of the detectors, enabling key rates to exceed 100 Mbps~\cite{li2023high}. In contrast, KEM algorithms are restricted by the speed of classical processors, as they require thousands of operations to produce each key bit~\cite{kyber}. However, without classical relays or quantum repeaters, the performance of QKD rapidly declines as the communication distance increases, making it less effective than KEM over long distances.

To leverage the strengths of both protocols, we consider a scenario where two users, Alice and Bob, are separated by long distances, as shown in Fig.~\ref{fig:series}b. Instead of directly performing QKD or KEM, they first distribute key bits, $k_1$ and $k_3$, through QKD links with nearby data centers that are equipped with high-performance supercomputers. The two data centers then perform KEM to distribute key bits $k_2$, and they announce the XOR-ed results $k_{12} = k_1 \oplus k_2$ and $k_{23} = k_2 \oplus k_3$ to Bob. As the data centers are a centralized resource, they can be equipped with multiple KEM channels and high-performance or purpose-built hardware to enable symmetric key generation rates using KEM that significantly exceed those achievable with QKD.
Alice and Bob can then share the key bits $k_1$ with performance limited only by the lowest QKD key generation rate. As shown in Fig.~\ref{fig:series}a, when the longest QKD link is $10$ km, the overall key generation speed using commercial devices is approximately seven times higher than that of standalone KEM communication. This can be further improved with the use of state-of-the-art QKD devices.

Another motivation for combining KEM and QKD is to achieve a higher level of security. Both bare KEM and QKD protocols have security vulnerabilities when deployed in practice: KEM may eventually be broken algorithmically, posing a risk of becoming unreliable in the future, while current implementations of QKD may be susceptible to physical attacks due to device imperfections~\cite{xu2020secure}. Additionally, in scenarios where a QKD link involves classical relay nodes or the series-connection protocol discussed earlier, it becomes necessary to assume the trustworthiness of all intermediate nodes. In the series-connection protocol, if Eve manages to compromise either the QKD or KEM link, she could potentially access the final key bits shared by the users.
To address this, we explore how end users can enhance the security of distributed key bits when utilizing multiple key distribution channels. Specifically, we propose two parallel key-distribution designs: the XOR scheme and the secret-sharing (SS) scheme.

\begin{figure*}[t]
    \includegraphics[width=0.9\textwidth]{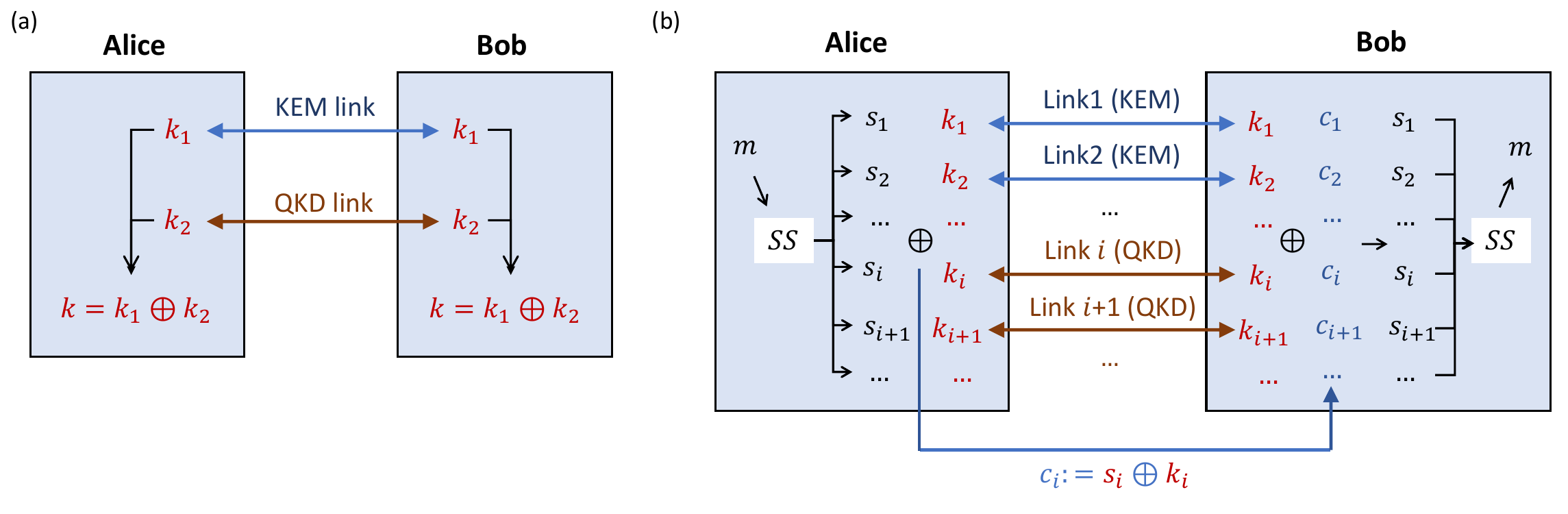}
    \caption{Parallel key distribution protocols. (a) Simple XOR protocol. Symmetric keys $k_1$ and $k_2$ generated via KEM and QKD respectively are combined into $k = k_1 \oplus k_2$. (b) Secret-sharing-based protocol. Alice encodes a random message $m$ to multiple shares $s_i$ with a secret sharing (SS) scheme, then distribute the shares by different links with Bob. They keep the final message $m$ as the secret key bits.} \label{fig:parallel}
\end{figure*}

In the simple XOR scheme shown in Fig.~\ref{fig:parallel}(a), Alice and Bob first use KEM and QKD as two separate channels to distribute key bits $k_1$ and $k_2$. The XOR of the two is then taken to generate the final shared key bits $k=k_1\oplus k_2$. As long as one of the input bits $k_1$ or $k_2$ is uniformly random, the output is uniformly random. Thus, Eve must learn both $k_1$ and $k_2$ in order to learn the shared key. We can also generalize the XOR scheme to the case where Alice and Bob own multiple parallel key distribution channels --- some of them are QKD links, while others are KEM --- by taking the XOR of all the key bits from each channel together to generate the final key. As before, Eve needs to break all the links to learn the shared key bits. However, the cost of generating the shared key for this protocol may be prohibitively large. Comparing the number of output bits to input bits for a protocol defines its information ratio $\eta$; for the XOR scheme, it is $\eta = 1/n$.  Additionally, in many scenarios we want to pursue a more complicated trust hierarchy --- some key distribution channels might be more trustworthy than others based on the particular implementation conditions.

To improve the symmetric key generation rate of the XOR scheme, we propose a key distribution method based on secret sharing. Here, we apply a variant of Shamir's secret sharing scheme~\cite{padro2012SS}. Suppose Alice wants to distribute a random message (i.e., secret) $m$ to Bob through $n$ key distribution channels. They aspire to achieve information-theoretic security for the secret $m$, even if some of the channels may be compromised. 
To this end, Alice utilizes polynomials over the finite field $\mr{GF}(q)$ for prime $q$,
\begin{equation}
    f=a_0 + a_1 x + a_2 x^2 + ... + a_{t-1} x^{t-1} \in \mr{GF}(q)[x],
\end{equation}
where the coefficients $a_0, a_1,..., a_{t-1}\in\mr{GF}(q)$ are chosen from the field.
Here, the rank of the polynomial defines the threshold $t$, so-called because the polynomial is uniquely determined by evaluating it for at least $t$ inputs. For our purposes, we can equate the threshold with the number of channels, $t=n$. Thus, the idea is to encode the secret in a privately-held polynomial, which is shared through $n$ evaluations.
The procedure is as follows. Alice first chooses $f$ by selecting the coefficients uniformly at random. She then chooses integer $g>1$ which determines the length of the secret $m$. 
Next, she determines a prime $q>n+g$ and selects $n+g$ different publicly-announced inputs $x_0, x_1, ..., x_{n+g-1}\in \mr{GF}(q)$; for instance, she can set $x_i = i$ for $i=0,1,...,n+g-1$. She announces the first $g$ inputs, and records the polynomial evaluations at these points as the secret $m :=(f(x_0),...,f(x_{g-1}))$.
The evaluations of the $n$ remaining inputs $(f(x_g), f(x_{g+1}),..., f(x_{n+g-1}))$ comprise the $n$ shares $s_1, s_2, ..., s_n$ of the polynomial. To securely distribute these shares to Bob, Alice then assigns the inputs to the $n$ independent KEM or QKD channels, and encrypts each share by consuming secret key bits $k_1, k_2, ..., k_n$ from the assigned channel. Finally, she announces the encrypted shares along with their corresponding inputs and channel assignments.
Upon receiving all the shares, Bob performs Lagrange interpolation to retrieve the coefficients $a_0, a_1, ..., a_{t-1}$ of $f$, from which he can then evaluate the secret $m$.

The above secret sharing scheme is information-theoretically secure: it can be proven that, when Eve can only learn at most $\Delta := t-g$ shares, she has insufficient information to restrict the possible values of the message and therefore cannot learn any information about the secret~\cite{padro2012SS}. In the whole scheme, we consume $n\log_2(q)$ symmetric key bits and distribute $g\log_2(q)$ bits of secret. The information ratio is
\begin{equation}
\eta = \frac{ g\log_2(q)}{ n\log_2(q) } = \frac{ n-\Delta }{n}.
\end{equation}
Consider the case when Alice and Bob hold five different KEM or QKD links. Suppose they want to ensure that Eve cannot obtain any information about $m$ when she breaks less than or equal to three of the channels, so they apply the secret sharing scheme with $n=t=5, g=2$. The information ratio is then $\eta=2/5$, which is higher than the use of the XOR scheme with four links whose $\eta$ is $1/4$.
If we set $\Delta$ to be a constant and increase $n$, the ratio $\eta$ will approach $1$, which implies that we can distribute almost the same amount of key bits as the naive usage of multiple channels while enhancing their security level.

In practice, we want to design the key distribution scheme with specific trust structures. For example, when the users share multiple KEM and QKD links, they might want to ensure that Eve can learn the final key bits only if she breaks at least one KEM and one QKD links. This can be guaranteed by introducing a more advanced secret sharing scheme with a specific access structure~\cite{Simmons1990how,Jackson1994,padro2012SS} $\mc{A}$ which is a set of subsets of all key distribution links. Only when Eve were to learn the shares $S$ distributed in the set $A\in\mc{A}$ can she learn all the secrets. For our purpose, suppose we have one KEM link $L_1$ and three QKD links $L_2, L_3, L_4$, we can set the access structure $\mc{A}$ to be $\mc{A}_m := \{\{L_1, L_2\}, \{L_1, L_3\}, \{L_1, L_4\}\}$ and all the combinations of links containing  one of the sets in $\mc{A}_m$. 
A secret sharing scheme with these requirements can be designed using linear codes~\cite{brickell1990some,Benaloh1990generalized}. We introduce linear code secret sharing in detail and design a linear code secret sharing scheme for the above access structures in the supplementary materials~\cite{supplementary}.
To boost the performance of linear-code secret sharing, one can use multi-linear secret sharing schemes~\cite{bertilsson1993construction,vanDijk1995linear,beimel2014multi}.
The secret-sharing-based key distribution scheme can also provide some other practical advantages. For example, the users may want to verify the correctness of the shared secrets without revealing them. This can be done with verifiable secret sharing~\cite{Chor1985verifiable}.  

In a realistic hybrid quantum network, information may need to travel through intermediary links and nodes before it reaches its destination. To assess the security of a protocol implemented in such a network one must consider the vulnerability of all the involved links and nodes. For example, consider replacing a long QKD link between Alice and Bob with many short QKD links in series with intermediate trusted nodes. Since the links are shorter, the overall key rate is higher, but the trusted nodes can increase the vulnerability of the protocol. 
In such a scheme, if even one of the nodes is compromised, then the secret key can be fully exposed. Vyas and Mendes~\cite{vyas2024relaxingtrustassumptionsquantum} suggest a protocol to relax the trust requirement by connecting each node to a highly trusted central key management system (KMS). 
Instead of using the generated QKD keys to decrypt and re-encrypt information, each node now takes the XOR of the keys distributed over its two adjacent links to generate a mask, which is then submitted to the KMS. 
In this scheme, no single node except for Alice and Bob ever has enough information to learn the secret key, assuming communication with the KMS is secured (for example, through KEM). 
To learn the secret key, an attacker would need to learn not only a raw QKD key, but also all of the masks (by compromising the KMS node or its communication links).

More generally, the security of hybrid protocols that build upon the series and parallel combination of QKD and KEM can be examined by representing a hybrid network as a graph $\mathcal{G} = (\mathcal{N}, \mathcal{E})$, where $\mathcal{N}$ is the set of nodes and $\mathcal{E}$ is the set of QKD and KEM links between the nodes. We allow for $\mathcal{G}$ to have multiple edges between the same pair of nodes. This corresponds, for example, to the presence of both a QKD link and a KEM link between two nodes. A key sharing protocol $P_{A,B}$ between nodes $A$ and $B$ can be constructed from sub-protocols combined in series or in parallel, using the techniques described earlier. 
This protocol can be compromised by attacking some subset of the edges or nodes it uses. Formally, a vulnerability of a protocol is a set $v\in2^{\mathcal{N}\cup\mathcal{E}}$ of network elements that, if attacked as a unit, would expose the shared key. We define the total vulnerability set of a protocol $V_\text{tot}(P_{A,B})$ as the set containing all possible vulnerabilities for protocol $P_{A,B}$. The minimal vulnerability set $V_\text{min}(P_{A,B})$ is the subset of $V_\text{tot}(P_{A,B})$ containing all of the vulnerabilities that have no strict subset in $V_\text{tot}(P_{A,B})$, thus describing the smallest set containing all the units that Eve could choose to attack in order to get the final key.

Simple rules can be used to construct the minimal vulnerability set of a composite protocol from the vulnerabilities of the sub-protocols that comprise it. To assess the security of different protocols enabled by the network, one can define a security function which assigns a value to a protocol's minimal vulnerability set. Similarly, rules can be defined to calculate the key generation rate of a hybrid protocol, thereby quantifying its performance. Details of the rules for constructing vulnerability sets and assessing key generation rates, as well as a formal mathematical description of a protocol can be found in the supplementary material~\cite{supplementary}. Users can then choose a protocol based on application-specific criteria. For instance, users requiring fast communication might use the fastest protocol that still achieves some minimum accepted security value. 

Future developments in real-world hybrid networks will require considerations of the allocation of shared networks resources. Utility functions for this purpose have been extensively explored for classical networks, while their quantum equivalents are actively being studied~\cite{vardoyan2023,gauthier2024}. For example, overall performance optimization of a key distribution network while still meeting desired security requirements could be achieved using techniques similar to those in Zhou et al.~\cite{zhou2022quantum}, where Lyapunov optimization is applied to maximize utility in QKD networks by designing an efficient key management and data scheduling algorithm. Likewise, a key management algorithm for the hybrid quantum-classical network that dynamically balances key generation and consumption can be defined.

A deployed hybrid network must also carefully consider granular security details in the distribution, combination, and application of the keys to ensure smooth operation even after a potential security breach. The Muckle protocol proposed by Dowling et al.~\cite{dowling2020many}, which implements a parallel protocol similar to the XOR scheme, provides a framework to understand its security against adversaries. Moreover, the protocol possesses desirable qualities such as forward security and post-compromise security. Garms et al. \cite{garms2024experimental} experimentally demonstrate a modified version of this protocol which exactly implements an XOR scheme to ensure the shared key will retain information-theoretic security. An extension of their techniques to the secret-sharing scheme and series combination will be critical for future application of our work. Additionally, we can use our analysis tools to study the Muckle protocols by defining a custom security function that encapsulates the security awarded by authentication, assigning greater security to a ``Muckle link'' than a naive XOR link. More generally, one can abstract away the security details of arbitrary hybrid protocols with custom security functions, using the vulnerability sets for the applicable network structure to calculate overall security. Thus, our work provides a new direction for designing hybrid quantum-classical networks for secure communications and cryptographic systems.

\begin{acknowledgments}

We thank Jens Eisert, Yi Li, Mouktik Raha, Grant Smith, Han Zheng, and Changchun Zhong for fruitful discussions. P.Z., D.B., and L.J. acknowledge support from the ARO (W911NF-23-1-0077), ARO MURI (W911NF-21-1-0325), AFOSR MURI (FA9550-19-1-0399, FA9550-21-1-0209, FA9550-23-1-0338), DARPA (HR0011-24-9-0359, HR0011-24-9-0361), NSF (OMA-1936118, ERC-1941583, OMA-2137642, OSI-2326767, CCF-2312755), NTT Research, Packard Foundation (2020-71479), and the Marshall and Arlene Bennett Family Research Program. F.R. likewise acknowledges support from the NSF CQN (ERC-1941583). P.Z., D.B., and L.J. further acknowledge that this material is based upon work supported by the U.S. Department of Energy, Office of Science, National Quantum Information Science Research Centers and Advanced Scientific Computing Research (ASCR) program under contract number DE-AC02-06CH11357 as part of the InterQnet quantum networking project. J.A.M.M., N.B., A.K., M.T.S., T.Z., F.J.H., and D.D.A. acknowledge additional support provided by Q-NEXT, part of the U.S. Department of Energy, Office of Science, National Quantum Information Science Research Centers, and the AFOSR MURI (FA9550-23-1-0330). J.L. acknowledges startup funds provided by the University of Pittsburgh and funding from IBM Quantum through the Chicago Quantum Exchange.

\end{acknowledgments}

\bibliographystyle{apsrev4-1}
\bibliography{bibPQCQKD.bib}

\end{document}



\title{Supplementary Material for ``Practical hybrid PQC-QKD protocols with enhanced security and performance''}

\maketitle

\section{Performance and Security Analysis of Hybrid Protocols} \label{sec:perf_rules}
\subsection{Formal Definition of a Hybrid Protocol}
\begin{figure*}[t]
    \includegraphics[width=1\textwidth]{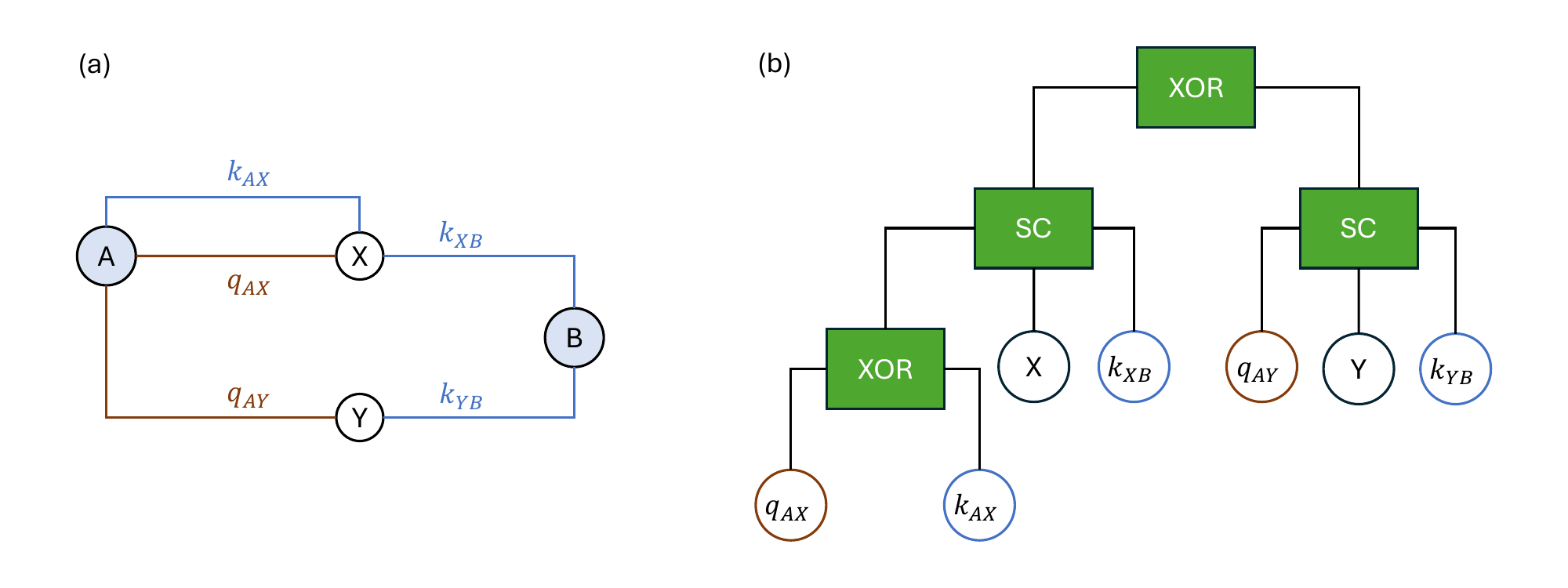}
    \caption{(a) An example of a hybrid PQC-QKD network with user nodes A and B and intermediate nodes X and Y. The blue edges represent KEM links, while the brown edges represent QKD links. (b) Example protocol tree for protocol $P_{AB}$, the XOR combination of $P_{AXB}$ and $P_{AYB}$. Protocol $P_{AXB}$ is a series combination of an XOR protocol $P_{AX}$ and KEM through $k_{XB}$. Protocol $P_{AYB}$ is a simple series combination (SC) of QKD through link $q_{AY}$ and KEM through link $k_{YB}$. Protocol $P_{AX}$ is just the XOR combination of QKD via $q_{AX}$ and $k_{AX}$.}
    \label{fig:network}
\end{figure*}

Formally, the protocols used in this paper can be defined by a tree structure. Each node of the tree can be either a protocol for combining keys, or an element of the network graph $\mc{G}$. As an example of a realistic hybrid network, consider the network graph depicted in figure \ref{fig:network}. \textit{A} can share a secret key with \textit{B} via node \textit{Y}: the QKD protocol $P_{AY}$ between \textit{A} and \textit{Y} gets combined in series with the KEM protocol $P_{YB}$ between \textit{Y} and \textit{B}, Resulting in the protocol $P_{AYB}$. Alternatively, \textit{A} can share a key with \textit{B} via node \textit{X}: QKD protocol $P_{AX}^\text{QKD}$ and kEM protocol $P_{AX}^\text{KEM}$, both between \textit{A} and \textit{X}, get combined in parallel, resulting in protocol $P_{AX}$, and this protocol gets combined in series with the KEM protocol $P_{XB}$, resulting in a protocol $P_{AXB}$. Finally, the protocols $P_{AYB}$ and $P_{AXB}$ can get combined together in parallel, resulting in a protocol $P_{AB}$. This example is illustrated by the tree in figure \ref{fig:network}.

\subsection{Secret Key Generation Rate of a Hybrid Protocol}
To assess the performance of a protocol, we begin by assigning a key distribution rate $K$ to each link in the network. For QKD link $q$, this rate takes into account the physical parameters of the sources, modulators, and detectors, as well as transmission loss. In particular, this value depends on the transmission distance. For PQC link $k$, this rate accounts for the various computational steps in generating a key, including encapsulation and decapsulation, and is largely independent of distance. For simplicity, we assume that operations to combine key (such as XOR and polynomial secret sharing), as well as the cost of classical communication, are negligible. We also assume that each node is not limited by the size of its buffer.

For a protocol $P$, represented by its tree structure, each leaf node corresponding to a link now has an associated rate. We prescribe the following simple rules to calculate the performance:
\begin{enumerate}
    \item A protocol $P_{A,B}$ between nodes $A$ and $B$ involving only edges $e_i$ with corresponding rates $K_i$, whose keys are \text{NOT} combined (such as by XOR or secret sharing) has an overall rate \[ K_{A,B} = \sum_{i} K_i.\]
    \item A protocol $P$ comprised of two sub-protocols $P_1$ and $P_2$ with respective rates $K_1$ and $K_2$ combined in series has an overall rate
    \[K^\text{series}(P) = \min(K_1,K_2).\] More generally, if protocols $\{P_1, \ldots, P_n\}$ with respective rates $\{K_1, \ldots, K_n\}$ are combined in series, the overall rate is \[K^\text{series}(P) = \min_i K_i.\]
    \item A protocol $P$ comprised of a set of sub-protocols $\{P_1, \ldots, P_n\}$ with rates $\{K_1, \ldots, K_n\}$ combined in parallel via XOR has a rate
    \[K^\text{XOR}(P) = \eta^\text{XOR}\cdot n\cdot \min_i K_i = \min_i K_i.\] Here, $\eta$ refers to the information ratio defined for the parallel protocols in the main text. 
    \item The secret sharing scheme $S$ described in the main text is constructed from a set of protocols $\{P_1,\ldots P_n\}$ combined in parallel with specified key length $g$. If the sub-protocols have associated rates $\{K_1, \ldots, K_n\}$, then the overall scheme has rate
    \[K^\text{SS}(S) = \eta^{SS} \cdot n \cdot \min_i K_i = g \min_i K_i.\]
\end{enumerate}

Thus, for any protocol involving these combinations, the end-to-end generation rate can be straightforwardly determined. These rules can be modified to incorporate more complex combinations (see \ref{sec:KMS_disc} and \ref{sec:SS}), or situations in which performing the combination has a significant computational overhead.

\subsection{Vulnerability Sets of a Hybrid Protocol} \label{sec:sec_rules}

In the main text we define the minimal and total vulnerability sets, $V_\text{min}$ and $V_\text{tot}$. More generally, we define a vulnerability set as any set $V$ that satisfies $V_{\min} \subseteq V \subseteq V_\text{tot}$. Having decomposed a key distribution protocol as a tree structure shown in Fig.~\ref{fig:network}, we can then construct vulnerability sets for hybrid protocols using the following rules:
\begin{enumerate}
    \item A protocol $P_{A,B}$ between nodes $A$ and $B$ involving only a set of edges $E$ (i.e., no intermediate nodes) has a vulnerability set 
    \[V(P_{A,B}) = \big\{\{e \mid e \in E\}\big\}.\]
    \item A protocol $P$ comprised of two sub-protocols $P_1$ and $P_2$ combined in series via a node $M$ has a vulnerability set 
    \[V^\text{series}(P) = V_\text{min}(P_1) \cup \{\{M\}\} \cup V_\text{min}(P_2).\]
    \item A protocol $P$ comprised of a set of sub-protocols $\{P_1, \ldots, P_n\}$ combined in parallel via XOR has a vulnerability set
    \[V^\text{XOR}(P) = \left\{\bigcup_{i=1}^n v_i \mid v_i \in V_\text{min}(P_i) \right\}\]
    \item A secret sharing scheme $S$ can be constructed from a set of protocols $P = \{P_1,\ldots P_n\}$ with an access structure $\mathcal{A} \subseteq 2^P$. A vulnerability set of this scheme is given by
    \[V^\text{SS}(S) = \bigcup_{A\in\mathcal{A}} V^\text{XOR}(A) \]
\end{enumerate}
These rules provide a vulnerability set which may not be minimal. One can always purge elements of a vulnerability set that are supersets of other vulnerabilities in the set to find the minimal vulnerability set of the protocol.

As an example, consider once again the hybrid protocol described by the tree in Figure~\ref{fig:network}b. The QKD protocol $P_{AY}$ between \textit{A} and \textit{Y} has a minimal vulnerability set $V_\text{min}(P_{AY}) = \{\{q_{AY}\}\}$. The KEM protocol $P_{YB}$ between \textit{Y} and \textit{B} has minimal  vulnerability set $V_\text{min}(P_{YB}) = \{ \{k_{YB} \} \}$. The overall vulnerability set of $P_{AY}$ and $P_{YB}$ combined in series is $V_\text{min} (P_{AYB}) = \{ \{q_{AY}\}, \{Y\}, \{k_{YB} \} \}$. An alternative way to share a key between \textit{A} and \textit{C} is through node \textit{X}. QKD and KEM between A and X could be used in parallel, combined via XOR, resulting in a protocol $P_{AX}$ with minimal vulnerability set $V_\text{min}(P_{AX}) = \{ \{k_{AX}, q_{AX}\} \}$ . The KEM protocol $P_{XB}$ between \textit{X} and \textit{B} has minimal vulnerability set $V_\text{min} = \{ \{k_{XB}\} \}$. The overall vulnerability of $P_{AX}$ and $P_{XB}$ combined in series is $V_\text{min}(P_{AXB}) = \{ \{k_{AX}, q_{AX}\}, \{X\}, \{k_{XB}\}\}$. To enhance security, \textit{A} might choose to use both protocols $P_{AXB}$ and $P_{AYB}$ in parallel, combining the keys via XOR into protocol $P_{AB}$. In this way, Eve would need to compromise at least one vulnerability in each protocol in order to learn the key. The minimal vulnerability set of this system is
\begin{align*}
    V_\text{min}(P_{AB}) = \{ &\{k_{AX}, q_{AX}, q_{AY}\}, \{k_{AX}, q_{AX}, Y\}, \{k_{AX}, q_{AX}, k_{YB}\}, \\ &\{X, q_{AY}\}, \{X, Y\}, \{X, k_{YB}\}, \\ &\{k_{XB}, q_{AY}\}, \{k_{XB}, Y\}, \{k_{XB}, k_{YB}\}\}.
\end{align*}

\begin{figure}[t]
    \centering
    \includegraphics[width=0.8\linewidth]{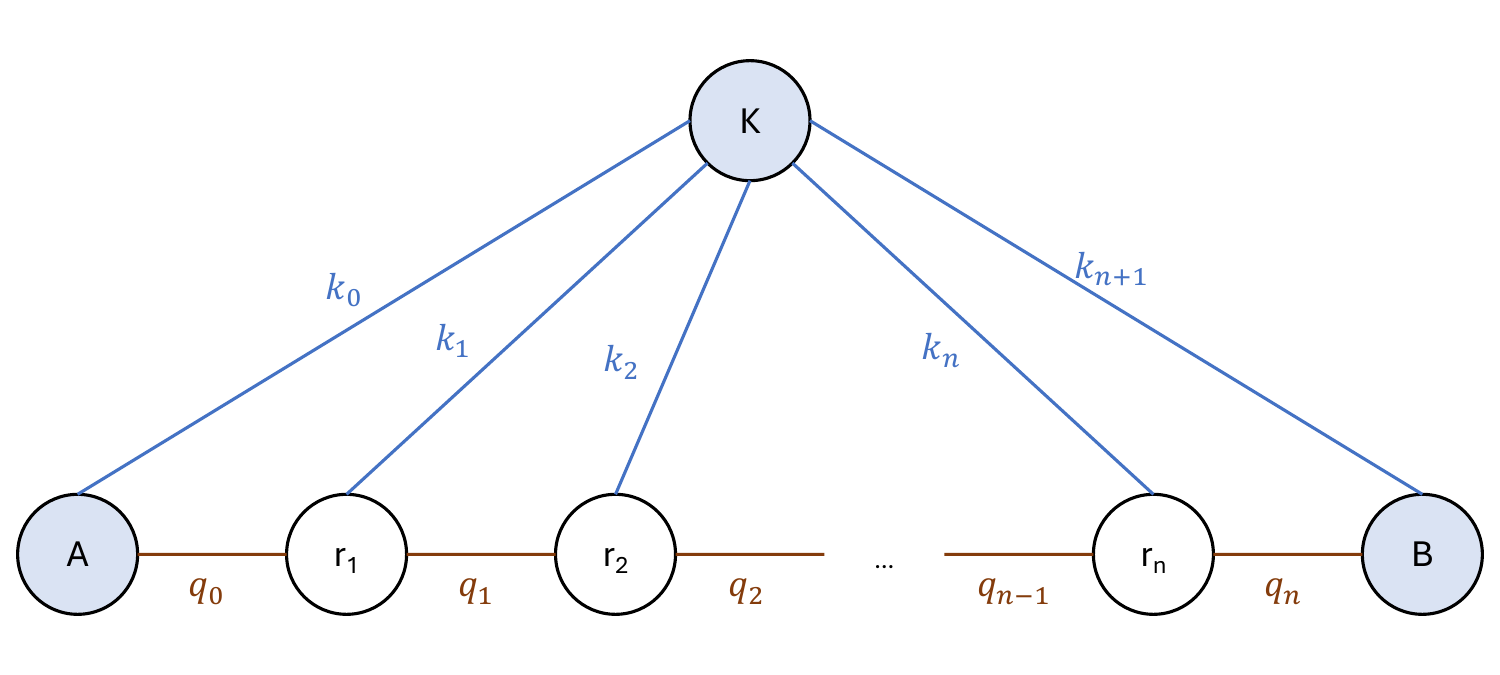}
    \caption{Key Management System (KMS) used to increase security of point-to-point QKD protocol, as proposed by Vyas and Mendes~\cite{vyas2024relaxingtrustassumptionsquantum}}
    \label{fig:kms}
\end{figure}

\subsection{Compatibility with Unexplored Combinations}\label{sec:KMS_disc} While our work does not cover all possible hybrid combinations, our framework\textemdash using a tree description of a protocol to perform a nested key rate calculation and vulnerability set enumeration\textemdash can generally be applied with a modification of the defined rules. Consider the Key Management System (KMS) described by Vyas and Mendes \cite{vyas2024relaxingtrustassumptionsquantum}, depicted in figure \ref{fig:kms}. We briefly review this design and, as an illustrative example, analyze its security in our terms.

In this scheme, $n$ relay nodes $\{r_1, \ldots, r_n\}$ connect Alice ($A$) and Bob ($B$) via QKD links $\{q_0, q_1,\ldots, q_n\}$, while KEM links $k_i$ connect each relay node $r_i$ to the central node $K$. Moreover, Alice (Bob) shares a KEM link $k_0$ ($k_{n+1}$) with the central node. In this protocol, no intermediate node obtains enough information to learn the final shared key. To achieve this, each relay node $r_i$ performs an XOR on the keys generated via adjacent quantum links $q_{i-1}$ and $q_i$ and submits the results to the central KMS through KEM link $k_i$, announcing $m_i = q_{i-1} \oplus q_i \oplus k_i$. Alice generates random key bits $s$\textemdash intended as the final shared key\textemdash and sends them to the central node via $k_0$, masked by the QKD key generated on her link $q_0$. In other words, she announces $m_0 = s \oplus q_0 \oplus k_0$. On the other end, Bob does not send any information. Instead, the central KMS sequentially performs an XOR on all the masked key bits along with their respective KEM keys $c = \oplus_{i=0}^n \left(m_i \oplus k_i\right) = s \oplus q_n$ to acquire the shared key $s$ masked with the key generated on Bob's link $q_n$. This result is then sent to Bob to be decrypted through $k_{n+1}$ as $m_n = c \oplus k_{n+1} = s \oplus q_n \oplus k_{n+1}$. The KMS scheme thus distributes $s$ from Alice to Bob, whereas no intermediate node or eavesdropper witnesses the raw key.

We can describe the vulnerabilities of the scheme in different scenarios. Eve can obtain the final key by controlling the central node and a single quantum key. She could get access to a quantum key by compromising a single quantum link, $V_1 = \{\{K, q_i\} \}_{i =0}^n$, or a single relay node,  $V_2 = \{\{K, r_i\}\}_{i =0}^n$. Eve could, alternatively, compromise a single quantum key and then obtain all the masks before or after it to backtrack the symmetric key. This can be done by compromising a quantum link, $V_3 = \{\{q_i,k_0, \ldots, k_i\}\}_{i=0}^n \cup \{\{q_i,k_{i+1}, \ldots, k_{n+1}\} \}_{i=0}^n$, or a relay node, $V_4 = \{\{r_i,k_0, \ldots, k_i\}\}_{i=1}^n \cup \{\{r_i,k_{i+1}, \ldots, k_{n+1}\} \}_{i=1}^n$. The overall vulnerability set of this protocol is the union of all these possibilities:
\[V = V_1 \cup V_2 \cup V_3 \cup V_4.\]

\section{Performance simulation details}

In this section, we describe how figure 2a of the main text is generated. First, we specify realistic parameters for QKD and PQC key generation rates. Then, we employ the rules in \ref{sec:perf_rules} to evaluate the overall key generation rate for the network employing a series connection in main text figure 2b. The example demonstrates a potential performance enhancement afforded by the series connection.

\subsection{QKD simulation formulas and parameter choices}

We consider the asymptotic key rate simulation of the decoy-state BB84 protocol~\cite{li2023high}. The key generation speed formula is
\begin{equation}
    K_{QKD} = \mr{CR}\cdot (P_z^2) P_\mu Q_\mu r,
\end{equation}
where $\mr{CR}$ is the clock rate of the system, $P_z$ is the probability where Alice and Bob choose $Z$-basis, $P_\mu$ is the probability of Alice to choose the signal intensity $\mu$, $Q_\mu$ is the gain, i.e., the probability that the signal state generates a successful detection on Bob's side, and $r$ is the key rate under detection. We have
\begin{equation}
\begin{aligned}
    Q_\mu &= 1 - (1-2p_d) e^{-\mu \eta}, \\
    r &= - f H(e_Z) + q_1 ( 1 - H(e_1) ),
\end{aligned} 
\end{equation}
where $p_d$ is the dark count probability, $\eta$ is the overall system transmittance, $\mu$ is the signal intensity, $f$ is the error correction efficiency, $e_Z$ is the quantum bit error rate of the $Z$-basis signal states, $q_1$ is the fraction of single-photon state among all the detected signals and $e_1$ is the phase error rate of the single-photon state. We have
\begin{equation}
\begin{aligned}
    q_1 &= Y_1 \frac{\mu e^{-\mu}}{Q_{\mu}}, \\
    e_1 &= e_X + (e_0 - e_X)\frac{Y_0}{Y_1}, \\
    Y_0 &= 2p_d, \quad Y_1 = 1- (1-2 p_d)(1-\eta),
\end{aligned}
\end{equation}
where $e_0=0.5$ is the error rate of vacuum state, $e_X$ is the error rate of $X$-basis signal state, $Y_0$ and $Y_1$ are, respectively, the probability of successful detection of the vacuum and single-photon states. The simulation formulas are from Ref.~\cite{ma2005practical}.

For the simulation of commercial and state-of-the-art QKD systems, we consider different parameter choices based on Ref.~\cite{yuan2018T12} and Ref.~\cite{li2023high}, respectively. The only different data is that we set the dark count rate $p_d=1\times 10^{-6}$ in the commercial QKD key rate simulation ($p_d=4.5\times 10^{-4}$ in Ref.~\cite{yuan2018T12}), which is easily achievable with today's commercial devices. This will not affect the short distance (i.e., $<50$ km) performance, but will lead to a longer communication distance.
We list the parameters in Table~\ref{tab:QKDpara}.

\begin{table}[h]
    \centering
    \begin{tabular}{|c|c|c|}
        \hline
        Parameters & Commerical QKD~\cite{yuan2018T12} & State-of-the-art QKD~\cite{li2023high}  \\
        \hline
        Clock rate $\mr{CR}$ & 1 GHz &  2.5 GHz\\
        \hline
        $Z$-basis ratio $P_z$ & 0.9668*(1-1/128) &  0.955 \\
        \hline
        Signal intensity ratio $P_\mu$ & 0.9697 & 0.88 \\
        \hline
        Detection efficiency & 0.31 & 0.56 \\
        \hline
        Dark count rate $p_d$ & $1\times 10^{-6}$ (5000 counts each second)  & $1\times 10^{-8}$ (50 counts each second) \\
        \hline
        Quantum bit error rate $e_Z$  & $3\%$ & $0.5\%$ \\
        \hline
        $X$-basis error rate $e_X$ & $3\%$ & $4\%$ \\
        \hline
        Error correction efficiency $f$ & 1.3 & 1.04 \\
        \hline
        Signal intensity $\mu$ & 0.4 &  0.54 \\
        \hline
    \end{tabular}
    \caption{The parameters we applied for the QKD performance estimation.}
    \label{tab:QKDpara}
\end{table}

\subsection{PQC simulation details}

For the estimation of key generation speed of Kyber-1024 with personal computer, we consider the benchmark data on the official website of Kyber~\cite{kyber}. 
We assume the PC perform the algorithm with an optimized implementation using AVX2 vector instructions. The Haswell cycles of key generation, encapsulation and decapsulation are $73544$, $97324$ and $79128$, respectively. After each round of KEM, Alice and Bob will share $256$ key bits.
Suppose the clock rate of a typical PC is $3.0$ GHz, the key generation speed is
\begin{equation}
    K_{KEM} (\mr{bps}) = 256 \; t_{KEM}^{-1}; \quad t_{KEM} (s) = (73544+97324+79128)/(3e9).
\end{equation}

\subsection{Series combination simulation details}

To find the rate of the scenario described in figure 2b of the main text\textemdash consisting of users communicating via intermediate data centers\textemdash, we use the steps outlined in \ref{sec:perf_rules}. We consider the QKD links to be 10 km from their data centers, resulting in a rate of 21.23Mbps. The data centers provide custom hardware or additional processing power (i.e. they can produce many KEM keys in parallel, see rule 1 of \ref{sec:perf_rules}), and thus, we assume that they can generate secret key with each other at a rate above 21.23Mbps. Thus, combining these links in series results in an overall rate of 21.23Mbps. This rate is independent of the distance between Alice and Bob, and is faster than the direct PQC rate between them. 

\section{Linear-code-based secret sharing} \label{sec:SS}

One may want to generate protocols which have specific vulnerability structures. Here, we briefly review how to construct a secret sharing scheme based on linear codes and understand their security, building on rule 4 in~\ref{sec:sec_rules}. We mainly follow the approach in Ref.~\cite{massey1993minimal,yuan2006secret}.

Define the Hamming weight of a vector $v\in \mbb{F}_q^n$ as the total number of non-zero coordinates in $v$. An $[n,k,d;q]$ linear code $\mb{C}$ is a linear subspace of $\mbb{F}_q^n$ with dimension $k$ and minimum nonzero Hamming weight $d$. Denote the generator matrix of the code $\mb{C}$ as $G=(g_0, g_1, ..., g_{n-1}) \in \mbb{F}_q^{k\times n}$. The row space of $G$ will generate the whole code space of $\mb{C}$. The parity check matrix $H\in \mbb{F}_q^{(n-k)\times n}$ is defined so that the row space of $H$ is the null space of $G$.

In the secret sharing scheme based on $\mb{C}$, the secret $m$ is an element of $\mbb{F}_q$. The dealer first randomly chooses a vector $u=(u_0,...,u_{k-1})\in\mbb{F}_q^k$ such that $s=u\cdot g_0$. There are $q^{k-1}$ vectors satisfying this condition. The dealer then treats $u$ as an information vector and computes the corresponding codeword
\begin{equation}
    f = (f_0, f_1, ..., f_{n-1}) = u\cdot G \in \mbb{F}_q^{n},
\end{equation}
and distributes $f_i$ to participant $P_i$ as share for each $i\geq 1$.

Consider a vector $v = (v_0, v_1, ..., v_{n-1}) \in \mbb{F}_q^{n}$ in the row space of the parity check matrix $H$ such that $v_0 = 1$. We have
\begin{equation}
    v \cdot G^T = 0 \quad \Rightarrow \quad v \cdot f = 0.
\end{equation}
Recall that $f_0 = u g_0 = m$ is the message. If we denote the location of nonzero elements in $\{v_1, ..., v_{n-1}\}$ as $i_1, i_2, ..., i_t \in [n-1]$, we have
\begin{equation}
    m = v_{i_1} f_{i_1} + v_{i_2} f_{i_2} + ... + v_{i_t} f_{i_t}.
\end{equation}
Therefore, when the $i_1, i_2, ..., i_t$-th shares are gathered together, one can then retrieve the message. We have the following proposition.

\begin{proposition}[Access structure of linear-code secret sharing~\cite{massey1993minimal}] \label{prop:accessLC}
Consider an $[n,k,d;q]$ linear code $\mb{C}$ with the generator matrix $G=(g_0, g_1, ..., g_{n-1}) \in \mbb{F}_q^{k\times n}$ and parity check matrix $H\in \mbb{F}_q^{(n-k)\times n}$. Then in the perfect secret sharing scheme based on $\mb{C}$, a set of shares $\{f_{i_1},f_{i_2},...,f_{i_t}\}$ with $1\leq i_1 < ... < i_m\leq n-1$ and $1\leq t\leq n-1$ determine the secret $m$ if and only if there is a vector
\begin{equation}
    (1,0,...,0,v_{i_1},0,...,0,v_{i_t},0,...,0),
\end{equation}
in the row space of $H$, where $v_{i_j}\neq 0$ for at least one $j$.
\end{proposition}

A set of shares is referred to as a \emph{minimal access set} if they can recover the secret while any of its subset cannot. The \emph{support} of a vector $v\in\mbb{F}_q^n$ is defined to be $\mr{supp}(v)=\{0\geq i\geq n-1: v_i \neq 0\}$. A vector $w$ covers $v$ if $\mr{supp}(v)\subseteq \mr{supp}(w)$. 
If a nonzero vector $w$ only covers its scalar multiples but no other nonzero vectors, then it is called a \emph{minimal vector}.
From Proposition~\ref{prop:accessLC} we can see that, if we solve the minimal vectors of $rs(H)$ with the first coordinate to be $1$, we then solve the minimal access set problem.

If we can construct a $H$ where all the nonzero vectors in $rs(H)$ are minimal vectors, we can then easily determine the minimal access set of $\mc{C}$. We have the following proposition.
\begin{proposition} [Minimal access sets by minimal vectors in parity check matrix~\cite{yuan2006secret}]
Consider an $[n,k,d;q]$ linear code $\mb{C}$ with the generator matrix $G=(g_0, g_1, ..., g_{n-1}) \in \mbb{F}_q^{k\times n}$ and parity check matrix $H=(h_0, h_1, ..., h_{n-1})\in \mbb{F}_q^{(n-k)\times n}$. Denote $r:=n-k$. If each nonzero vector in $rs(H)$ is minimal, then in the secret sharing scheme based on $\mb{C}$, there are $q^{r-1}$ minimal access sets. In addition, we have
\begin{enumerate}
    \item If $h_i (i\in 1,...,n-1)$ is a scalar multiple of $h_0$, then the participant $P_i$ must be in every minimal access set. Such a participant is called a \emph{dictatorial participant}.
    \item If $h_i (i\in 1,...,n-1)$ is not a scalar multiple of $h_0$, then participant $P_i$ must be in $(q-1)q^{r-2}$ out of $q^{r-1}$ minimal access sets.
\end{enumerate}
\end{proposition}

For our purpose of PQC-QKD network, we numerically search and design a linear code define on $\mbb{F}_5$ with the generator matrix $G$ and $H$ given by
\begin{equation}
\begin{aligned}
G = 
\begin{bmatrix}
1 & 0 & 0 & 0 & 4 \\
0 & 1 & 1 & 2 & 3
\end{bmatrix}, \quad 
H = 
\begin{bmatrix}
0 & 2 & 3 & 0 & 0 \\
2 & 2 & 4 & 4 & 2 \\
3 & 3 & 0 & 4 & 3
\end{bmatrix}.
\end{aligned}
\end{equation}
This provides us a secret sharing scheme among 4 parties $P_1, P_2, P_3, P_4$.
Based on Proposition~\ref{prop:accessLC}, we can solve the minimal access structure of the code: $\{P_1, P_2\}, \{P_1, P_3\}, \{P_1, P_4\}$. Now we let $P_1$ to be the share distributed by the KEM link, while $P_2, P_3, P_4$ be the shares distributed by the QKD links. By the linear-code secret sharing scheme, only when the attackers break the KEM link and either QKD link at the same time can they learn the final shared key bits.

\bibliographystyle{apsrev}
\bibliography{bibPQCQKD} 